\documentclass[conference]{IEEEtran}
\IEEEoverridecommandlockouts
\usepackage{multirow}
\usepackage{boldline,multirow}
\usepackage{cite}
\usepackage{tikz}
\usepackage{algorithm}%
\usepackage{algorithmicx}%
\usepackage{algpseudocode}%
\usepackage{graphicx}
\usepackage{textcomp}
\usepackage{xcolor}
\usepackage{booktabs}
\usepackage{pgfplots}
\usepackage{ulem}
\usepackage{amsthm, amsmath,amssymb,amsfonts}%
\usepackage{bm}%
\usepackage{mathrsfs}
\newtheorem{example}{Example}
\usepackage{multirow}
\pgfplotsset{compat=1.18}

\DeclareSymbolFont{mathx}{U}{mathx}{m}{n}
\DeclareMathSymbol{\bigtimes}{1}{mathx}{"91}
\begin{document}

\title{Qimax: Efficient quantum simulation via GPU-accelerated extended stabilizer formalism}

\author{
\IEEEauthorblockN{Vu Tuan Hai\textsuperscript{1,2}, Bui Cao Doanh\textsuperscript{1,2}, Le Vu Trung Duong\textsuperscript{3}, Pham Hoai Luan\textsuperscript{3} and Yasuhiko Nakashima \textsuperscript{3}}
\IEEEauthorblockA{
\textsuperscript{1} University of Information Technology, Ho Chi Minh City, 700000, Vietnam.\\
\textsuperscript{2} Vietnam National University, Ho Chi Minh City, 700000, Vietnam.\\
\textsuperscript{3} Nara Institute of Science and Technology, 8916–5 Takayama-cho, 630-0192 Ikoma, Nara, Japan.
\\
Email: haivt@uit.edu.vn} 
}
\maketitle

\begin{abstract}
Simulating near-Clifford quantum circuits using the extended stabilizer formalism remains computationally challenging due to rapid stabilizer rank growth and limited parallel scalability. In this work, we present Qimax, a reformulation of the extended stabilizer framework designed to shift the practical boundary of classical quantum circuit simulation. Rather than applying gates sequentially to stabilizer generators, Qimax introduces an operator-level decomposition that restructures circuit execution into grouped single- and two-qubit operators. We further propose a GPU-native tensor encoding of stabilizer generators that enables parallel weight propagation and index-based Pauli accumulation. To mitigate the combinatorial explosion inherent in non-Clifford operations, we develop a sparse representation that preserves computational parallelism while substantially reducing memory overhead. This design transforms extended stabilizer simulation from a sequential procedure into a parallel model suitable for a GPU. Experimental results demonstrate that Qimax significantly extends the feasible simulation regime for deep near-Clifford circuits, achieving substantial speedups over other simulators while maintaining accuracy.
\end{abstract}


\section{Introduction}

Simulating a quantum system classically demands significant resources, scaling exponentially with the number of particles, posing a challenging computational problem. For instance, Google's quantum computer, Sycamore, demonstrated quantum supremacy by producing results in days that would take classical computers thousands of years to simulate \cite{sycamore, doi:10.1126/science.abn7293, Daley2022}. Various simulation methods have been proposed, including state-vector \cite{javadiabhari2024quantumcomputingqiskit, 10.21468/SciPostPhysCore.7.4.075}, tensor-network \cite{10313722}, decision diagram \cite{Vinkhuijzen2023limdddecision}, and stabilizer formalism \cite{stim}. Each method offers different levels of fidelity and resource requirements depending on the quantum circuit's size and type.

Quantum simulator development aims to address domain-specific and general-purpose applications, ultimately targeting the boundary of quantum advantage, where classical computers cannot simulate quantum systems within an acceptable execution time \cite{https://doi.org/10.1002/cpe.70628, https://doi.org/10.1002/cpe.70166, https://doi.org/10.1002/cpe.70332}. Simulator scalability is assessed by the number of qubits ($n$, or \#Qubits) and gates ($m$, or \#Gates). In the state-vector approach, the quantum state $|\psi(\bm\theta)\rangle$ is represented as a $2^n$-dimensional complex tensor, leading to an exponential increase in execution time and memory requirements with \#Qubits, while \#Gates affects execution time linearly. This exponential scaling is inherent to quantum systems in the state-vector approach. Recent stabilizer formalism research focuses on reducing gate application times for Clifford circuits.

Stabilizer formalism based on Heisenberg's picture \cite{Descamps_2024, Dauphinais2024stabilizerformalism} provides a compact representation for quantum systems and gate applications where the quantum system can be described as stabilizers. Quantum gates act on the stabilizers by mapping between Pauli strings rather than matrix-vector multiplication. Compared to the state-vector approach, the stabilizer formalism can solve the same problems with fewer computational resources, as it operates on stabilizers rather than full state vectors. Existing stabilizer formalism packages have been developed; although they can run a hundred-qubit Clifford circuit with a million gates, the stabilizer formalism limits the application of the quantum simulator, especially for algorithms that require non-Clifford gates. On the other hand, using non-Clifford gates in (extended) stabilizer formalism may lead to a fast increase in complexity. For instance, \cite{modelcounting} shows the exponential rise of stabilizer rank in cases of Quantum Neural Network (QNN), which require non-Clifford gates. Other studies address this by employing approximation methods, balancing complexity and accuracy \cite{Bravyi2019simulationofquantum}.

There are two use cases of stabilizer formalism: (1) large-qubit Clifford circuits, and (2) low-rank Clifford and non-Clifford. Because Stim \cite{stim} effectively handles the first, our research focuses on the second. Clifford circuits are used in many problems, such as: quantum error correction \cite{Acharya2023, Bravyi2024, Acharya2025}, randomized benchmarking \cite{PhysRevLett.123.060501, Polloreno2025theoryofdirect}, shadow tomography \cite{Akhtar2025dualunitaryshadow, 92ky-ln8f, 9pbp-jzr9} and magic state distillation \cite{thxx-njr6} . Near-Clifford circuits have wider applications. Our proposed package, Qimax, addresses this limitation by offering three modes: v1 (standard extended stabilizer formalism, CPU-based), v2 (parallel stabilizer formalism, GPU-based \cite{https://doi.org/10.1002/cpe.70328}, using one-hot vector Pauli encoding), and v3 (parallel stabilizer formalism, GPU-based, using tensor index encoding).

\section{Background}
\label{sec:background}

% \subsection{Pauli operators}

% We first introduce the Pauli operators and the Pauli group, which form the foundation of the stabilizer formalism. The single-qubit Pauli operators $p$ with matrix form (Pauli matrices) are $I,X,Y,Z$. An $n$-qubit Pauli string $P_n$ is a tensor product of single-qubit Pauli operators, $P_n=\{I, X, Y, Z\}^{\otimes n}$. The set of all such operators, together with phase factors $\{\pm 1, \pm i\}$, forms the $n$-qubit Pauli group $\{\mathcal{P}_n\}$. In the stabilizer formalism, $\{\mathcal{P}_n\}$ is normally called the stabilizer group.

% \begin{align}
% I =
% \begin{bmatrix}
% 1 & 0 \\
% 0 & 1
% \end{bmatrix},
% X =
% \begin{bmatrix}
% 0 & 1 \\
% 1 & 0
% \end{bmatrix},
% Y =
% \begin{bmatrix}
% 0 & -i \\
% i & 0
% \end{bmatrix},
% Z =
% \begin{bmatrix}
% 1 & 0 \\
% 0 & -1
% \end{bmatrix}.
% \end{align}

\subsection{Extended stabilizer formalism}

An $n$-qubit quantum state $|\psi\rangle$ is defined as a stabilizer state with eigenvalue $+1$ for every operator in its stabilizer group $\{\mathcal{P}_{n,0}, \mathcal{P}_{n,1}, \dots\}$, meaning $U|\psi\rangle=+1|\psi\rangle$. The stabilizer states form a strict subset of all quantum states that can be uniquely described by $\{\mathcal{P}_n\}$. The elements of the stabilizer group are called stabilizers. Any stabilizer group can be specified by a set of generators $G_{|\psi\rangle}\equiv G=\{\mathbb{P}_n\}$ so that every element $\mathcal{P}_{n,j}$ in the stabilizer group can be obtained through matrix multiplication between any pair $\{\mathbb{P}_{n, i}, \mathbb{P}_{n, j}\}$. The number of generators (order) of $G$ is always $n$, each generator $\mathbb{P}_{n,j}$ is also an $n$-qubit Pauli string. These generators evolve under the application of $m$ quantum gates, as shown in~\eqref{eq:generator}:

\begin{align}
    G^{(0)}=\left\langle\begin{array}{c}
    \mathbb{P}_{n, 0}^{(0)} \\
    \vdots \\
     \mathbb{P}_{n, n-1}^{(0)} \\
    \end{array}\right\rangle\xrightarrow{g^{m}} G^{(m)} =\left\langle\begin{array}{c}
    \mathbb{P}_{n, 0}^{(m)} \\
    \vdots \\
     \mathbb{P}_{n, n-1}^{(m)} \\
    \end{array}\right\rangle,\label{eq:generator}
\end{align}

% following by Table.~\ref{tab:additional_gate}, see Example.~\ref{example:basic}.

where $\mathbb{P}_{n, j}^{(0)}$ can be set as $\mathbb{Z}_{n,j}=I^{\otimes j}\otimes Z\otimes I^{\otimes (n-j-1)}$ and $G^{(0)}$ is the set of generator for initial state $|0\rangle^{\otimes n}$. Gate $g^{(t)}$ update $\mathbb{P}_{n, j}^{(t)}$ to $\mathbb{P}_{n, j}^{(t+1)}$. To achieve universal quantum computation, non-Clifford gates, such as rotation gates, must be included. These gates transform stabilizer states into non-stabilizer states. A generator may transform from a single Pauli string into a linear combination of Pauli strings $\mathbb{P}_n=\sum_j  \lambda_j P_{n,j}$ with $\lambda P_n=\lambda p_0\otimes \ldots \otimes p_{n-1},\;\equiv\lambda p_0\ldots p_{n-1}$ where $\lambda\in\mathbb{R}$ is the weight. We can relate $G^{(m)}$ directly to the target density operator either as a product of stabilizers or as a sum of Pauli operators, as shown in~\eqref{eq:density_operator2}, respectively: 

% \begin{example}
%     Consider a 3-qubit circuit $CX_{0,1}R_z(\pi/3)_0SX_0|000\rangle$. The evolution of the stabilizer generators $G$ is tracked gate by gate, as specified in Table~\ref{tab:additional_gate}. For instance, the Pauli string $(-Y)$ is converted to $(\frac{1}{2}X-\frac{\sqrt{3}}{2}Y)$:
%     \begin{equation}
%              \left\langle\begin{array}{c}
%             \mathbb{Z}_{3,0} \\
%             \mathbb{Z}_{3,1} \\
%             \mathbb{Z}_{3,2} \\
%             \end{array}\right\rangle \xrightarrow{SX_{0}}
%             \left\langle\begin{array}{c}
%             -YII \\
%             IZI \\
%             IIZ \\
%             \end{array}\right\rangle \xrightarrow{R_{z}(\frac{\pi}{3})_{0}}
%             \left\langle\begin{array}{c}
%             \frac{1}{2}XII-\frac{\sqrt{3}}{2}YII \\
%             IZI \\
%             IIZ \\
%             \end{array}\right\rangle \xrightarrow{CX_{0,1}}
%             \left\langle\begin{array}{c}
%             \frac{1}{2}XXI-\frac{\sqrt{3}}{2}YXI \\
%             ZZI \\
%             IIZ \\
%             \end{array}\right\rangle 
%     \end{equation}
%     \label{example:basic}
% \end{example}

\begin{align}
    \rho^{(m)}=\frac{1}{2^n}\prod_{j=1}^{n} \left(I^{\otimes n}+\mathbb{P}^{(m)}_{n,j}\right) 
    =\frac{1}{2^n} \sum_{\mathcal{P}_n \in \mathcal{P}_{|\psi\rangle}} \lambda_{\mathcal{P}_n}\mathcal{P}_n.
    \label{eq:density_operator2}
\end{align}

In the other cases, we can simply track $G$ to get measurement value as $p_{j,k} = \text{tr}([\frac{1}{2}(I^{\otimes n}+Z_k)]|\psi\rangle\langle\psi|)$. 

\subsection{Single-qubit Pauli measurement}
\label{sec:measurement}

Let $\mathbb{P}_{n,k,b} =\underbrace{ I \otimes \cdots \otimes |b\rangle\langle b|_k \otimes \cdots \otimes I}_{n \;\text{terms}} = \frac{1}{2}((-1)^{b}Z_{n,k} + I^{\otimes n})$ for $k \in [n]$. When measuring the $k^{\text{th}}$ qubit of an $n$-qubit state $|\psi\rangle$ using projectors $\{\mathbb{P}_{n,k,0}, \mathbb{P}_{n,k,1}\}$, we get two possible outcomes: 0 with probability $p_{k,0}$ and 1 with probability $p_{k,1}$. It follows that $p_{k,0} = \mathrm{Tr}(\mathbb{P}_{n,k,0} |\psi\rangle\langle\psi|)$ where $\mathrm{Tr}$ is the trace mapping. As shown in~\eqref{eq:density_operator2}, the density operator $|\psi\rangle\langle\psi|$ can be written using generators i.e., $|\psi\rangle\langle\psi| = \frac{1}{2^n} \sum_{\mathbb{P}\in G} \mathbb{P}$. The probability $p_{k,0}$ can be obtained as:

\begin{align}
p_{k,0} &= \text{Tr}(\mathbb{P}_{n,k,0} |\psi\rangle\langle\psi|) \nonumber\\
&= \text{Tr} \left( \frac{1}{2} (I^{\otimes n} + Z_{n,k}) |\psi\rangle\langle\psi| \right) \nonumber \\
&= \frac{1}{2} \left( \text{Tr}(|\psi\rangle\langle\psi|) + \text{Tr}(Z_{n,k} |\psi\rangle\langle\psi|) \right) \nonumber \\
&= \frac{1}{2} + \frac{1}{2^{n+1}} \sum_{\lambda_{\mathcal{P}_n} \mathcal{P}_n } \lambda_{\mathcal{P}_n}\text{Tr}(Z_{n,k} \mathcal{P}_n) \nonumber
\end{align}

For any Pauli string $P_n$, we can express $\mathrm{Tr}(P_n)$ as $\mathrm{Tr}(P_0) \cdots \mathrm{Tr}(P_{n-1})$. Since $\mathrm{Tr}(X) = \mathrm{Tr}(Y) = \mathrm{Tr}(Z) = 0$, $\mathrm{Tr}(P_n)\neq0$ if and only if $P_n = I^{\otimes n}$. Furthermore, $\mathbf{P} = Z_k$ if and only if $Z_kP_n = I^{\otimes n}$. Thus $p_k$ can be simplified as $\frac{1}{2} + \frac{1}{2} \sum_{\mathcal{P}_n = Z_k} \lambda_{\mathcal{P}_n}$.

\subsection{Limitation of extended stabilizer formalism}

The current generation of quantum computers, such as IBM Quantum Heron, IBM Quantum Eagle, Rigetti Anka-3, use a native gate set that includes $\{I, X, SX, R_z(\theta)/R_x(\theta), CZ/ISWAP\}$. There is only one non-Clifford gate that can build any other gate using the circuit transpiler. Qimax is not limited to these basic gates; additional gates can be supported with minor modifications to the create\_LUT() function. The Qimax's gate set used in this paper including $\{H,S,CX,R_x(\theta), R_y(\theta), R_z(\theta)\}$. The 3- and 4-qubit gates will be considered in the future version.

If a Clifford gate $g$ is applied to the stabilizer state, then $U \mathcal{P}_n U^{\dagger} g|\psi\rangle=g|\psi\rangle$. To be specific, a Clifford gate act on $j^{\text{th}}$ qubit (denoted as $g_j$) in the stabilizer $\mathcal{P}_n=\lambda\;p_0 \otimes \ldots \otimes p_{n-1}$, it return $g_j \mathcal{P}_n g_j^{\dagger}= \sum\lambda\;p_0 \otimes \ldots \otimes g_j p_j g_j^{\dagger} \otimes \ldots \otimes p_{n-1}$. Since only the $j^{\text{th}}$ entry needs to be updated, reducing $g_jp_j g_j^{\dagger}$ can be done in constant time.

\begin{figure}[ht]
    \centering
    \includegraphics[width=0.99\linewidth]{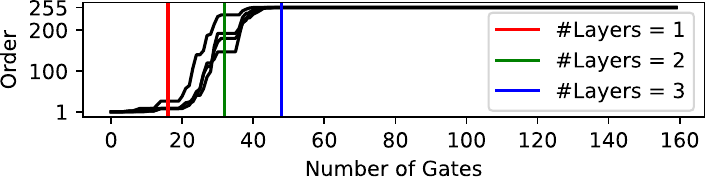}
    \caption{The order $n'$ on four generators is shown as four black lines, ranging from 1 to $4^4$. The gate application's order is from qubit $0^{\text{th}}$ to qubit $(n-1)^{\text{th}}$, left to right. We are considering the worst cases by using 4-qubit $|XYZ+\text{chain}\rangle$ ansatz, the circuit representation can be found in Figure.~\ref{fig:operator} (b).}
    \label{fig:rank}
\end{figure}

Applying non-Clifford gates increases the stabilizer rank of a stabilizer state $|\psi\rangle$, as generators evolve from single Pauli strings into linear combinations of Pauli strings. When non-Clifford gates act on every qubit, a generator initially represented as a single Pauli string becomes a sum of $n'$ Pauli strings, where $n'$ can grow to $4^n$. Subsequent gates require iterating over all $n'$ Pauli strings, making the computation significantly slower than the state-vector approach, which processes $2^n$-complex amplitudes. Figure~\ref{fig:rank} illustrates the increasing complexity of generators due to non-Clifford and two-qubit gate applications.

% \begin{example}
%     The generator $\mathbb{P}_{1}=Z$ turn to be $\cos(\theta_1)X-\sin(\theta_1)\cos(\theta_2)Z+\sin(\theta_1)\sin(\theta_2)Y$ after $R_y(\theta_1)$ and $R_x(\theta_2)$ actions.
%     \label{example:mapped_pauli}
% \end{example}

\section{Proposed techniques}
\label{sec:proposed}

This section presents the core architectural techniques of Qimax. Our approach restructures the simulation process to enable operator-level parallelism and tensor-based GPU execution. The following subsections describe the instruction grouping strategy, encoding architecture, and gate implementation methods.

\subsection{Instruction}

\begin{figure}
    \centering
    \includegraphics[width=0.99\linewidth]{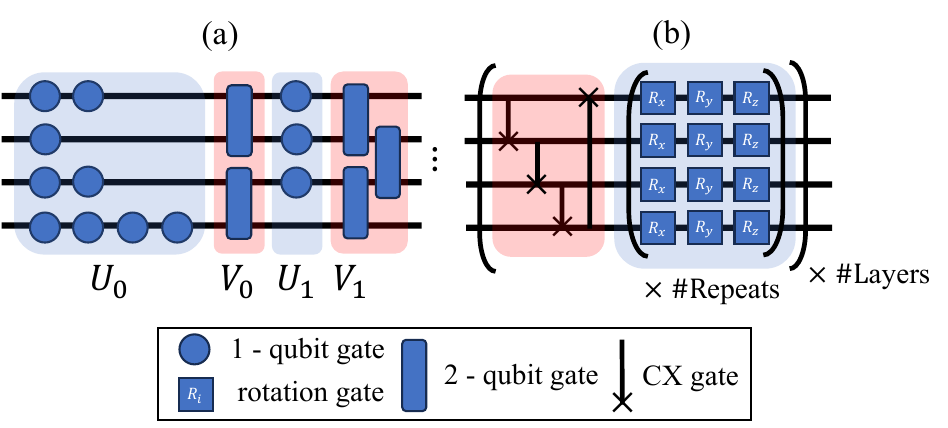}
    \caption{(a) A quantum circuit can be divided into $\{U_k\},\{V_k\}$, and end up with $U_{K-1}$ or $V_{K'-1}$ (b) $XYZ+W_{\text{chain}}$ topology.}
    \label{fig:operator}
\end{figure}

We present gates as instructions similar to Qiskit \cite{javadiabhari2024quantumcomputingqiskit}. Each instruction is a tuple $\{g, w, \theta\}$, representing the gate name, qubit (wire), and parameter value (if applicable), respectively. For a gate without parameters, $\theta$ is set to $0$.  Performing gate-by-gate on stabilizers leads to low performance; hence, we consider performing operators instead, where each operator contains gates of the same type. A quantum circuit is partitioned into $K$ $U_k$ (only 1-qubit gates) and $K'$ $V_k$ (only 2-qubit gates) because the 2-qubit gate has different behavior from other gates. These operators interleaved each other, as presented in Figure.~\ref{fig:operator}, obey $|K-K'|\leq 1$. The operator sequence is determined by the first instruction: if it is a single-qubit gate, the order will be $U_0, V_1, U_1, V_1,\ldots$; otherwise, it starts with $V_0$. Within each $U_k$, gates are further grouped by qubit, with
$U_{k,j}$ denoting instructions in the $k^{\text{th}}$ single-qubit operator acting on the $j^{\text{th}}$ qubit. By grouping gates, there is no difference between Clifford and non-Clifford gates.

Rather than applying gates individually to the stabilizers, Qimax groups gates into operators and applies these operators collectively. The primary computational cost arises from updating Pauli strings, which involves transforming the $j^{\text{th}}$ Pauli matrix in every Pauli string via conjugation. In the extended stabilizer formalism, applying $m$ gates individually yields a time complexity of $\mathcal{O}(mn')$, where $n'$ is the number of Pauli strings in the stabilizer representation. By grouping gates into $K$ single-qubit operators and $K'$ two-qubit operators, where $K+K'\ll m$, Qimax reduces the complexity to $\mathcal{O}((K+K')n')$. This reduction is effective because the number of operators is typically much smaller than \#Gates.

\subsection{Encoding}

We encode any Pauli string $P_n$ as an integer index using base-4 with the mapping $I=0, X=1, Y=2, Z=3$. The index ranges from $0$ (representing $I^{\otimes n}$) to $4^n-1$ ($Z^{\otimes n}$). Conversely, decoding an index back to a Pauli string requires \#Qubits. When a non-Clifford gate acts on a Pauli matrix
$p\neq I$ in $P_n$, it may transform into a linear combination $p=w_0X+w_1Y+w_2Z\equiv[w_0,w_1,w_2]$. We call $w_j$ the weight of transformed generators. Since every Pauli matrix $p_j$ in a generator may take this form, applying non-Clifford gates can transform a generator from a simple form $\mathbb{P}_n=\sum_{j}\lambda_j P_{n,j}$ (a sum of Pauli strings) to complex form $\hat{\mathbb{P}}_n=\sum_j\lambda_j\prod_{k=1}^{n}\left(\sum_l w_{k,l}p_{k,l}\right)$ ($p_{k,l}\in\{I, X, Y, Z\}$), where the product is over qubits and the sum is over Pauli matrices, as shown in Example.~\ref{example:expand_reduce}.

The weights for $\mathbb{P}_n$ are represented by a 3D tensor $\mathbf{w}$, where each entry $w\in\mathbf{w}$ satisfies $|w|\leq1$ due to the normalization condition. In Qimax v2, $\mathbf{w}$ is a fixed-size tensor padded with zeros, while in Qimax v3, it is a ragged tensor with corresponding indices $\mathbf{i}$ to store only non-zero weights. In the worst case, where the number of Pauli strings $n'=4^n$, the tensor $\mathbf{w}$ has dimensions 
$4^n\times n\times 4$, corresponding to $4^n$ Pauli strings, $n$ qubits and $4$ Pauli matrices ($I,X,Y,Z$). Qimax v2 maintains a consistent tensor size for all cases, simplifying parallel computation, whereas Qimax v3 reduces unnecessary data but doubles memory usage due to the storage of indices $\mathbf{i}$. On the other hand, parallelizing computation on Qimax v2 is better than v3. The indices $\mathbf{i}$ correspond to the coefficients $\bm{\lambda}$ in the simple form $\mathbb{P}_n = \sum_j \lambda_j P_{n,j}$ and have the same size as $\mathbf{w}$ in the complex form.

We denote:
$\Lambda=[\bm{\lambda}_0,\ldots,\bm{\lambda}_{n-1}], \bm{\lambda}_j\in \mathbb{R}^{n'},
\mathbf{W}=[\mathbf{w}_0,\ldots,\mathbf{w}_{n-1}],
    \mathbb{I}=[\mathbf{i}_0,\ldots,\mathbf{i}_{n-1}]
$ are the scalars, weights, and corresponding indices for the entire set of generators.

\begin{example}
    \label{example:expand_reduce}
    $\hat{\mathbb{P}}_2=2(3Y+4Z)Z + I(X+Z)$ is encoded as $\bm{\lambda}=[2,1]$ and \\
    $\mathbf{w} = [[[0,0,3,4],[0,0,0,1]],[[1,0,0,0],[0,1,0,1]]]$ using one-hot encoding  (Qimax v2). In the second approach, $\hat{\mathbb{P}}_2\triangleq \left\{ 
          \begin{array}{l}
            \bm{\lambda}=[2,1] \\
            \mathbf{w} = [[[3,4], [1]], [[1], [1,1]]] \\
            \mathbf{i}= [[[2,3],[3]], [[0], [1,3]]]\\
          \end{array}
        \right.$ (Qimax v3)
    . This complex form can be reduced to a simple form by multiple broadcasts between linear combinations. $\hat{\mathbb{P}}_2\xrightarrow{\text{flatten()}}\mathbb{P}_2=6YZ+8ZZ+IX+IZ\triangleq \left\{ 
          \begin{array}{l}
            \bm{\lambda}=[6,8,1,1] \\
            \mathbf{i}= [11,15,1,2]\\
          \end{array}
        \right.$. The detailed process is shown in~\ref{eq:flatten}.
\end{example}

\subsection{Gate}
\label{sec:gate}

\subsubsection{1-qubit gate}

Applying $m$ single-qubit gates to the generator set $G$ in a single step is more efficient than sequential application. This operation is notated as $\text{map}_{1q}: \mathbb{P}\times g^m\rightarrow \hat{\mathbb{P}}$ where $\mathbb{P}$ is a generator and $\hat{\mathbb{P}}$ is its transformed form. Because $\text{map}_{1q}$ only acts on $X, Y$ or $Z$ for every generator, we construct a look-up table called $\text{LUT}_{1q}$ to track the output of every $\text{map}_{1q}(w, U_{k,j})$. $\text{LUT}_{1q}$ is a tensor which has the shape $K\times n\times 3\times 3$. As shown in Algorithm.~\ref{algo:constructLUT}, each $\text{map}_{1q}(\ldots)$ is independent,  enabling parallel computation. Constructing the $\text{LUT}_{1q}$ can achieve a speedup of up to $K\times n\times 3$ times, depending on the number of available cores.

\begin{algorithm}
\caption{$\text{create\_LUT}_{1q}()$} 
\label{algo:constructLUT}
\begin{algorithmic}
\Require $\{U_{k,j}\}, K$
\State $\text{LUT}_{1q}\gets \bm 0^{K\times n\times3\times3}$
\For{$k$ in $[0\ldots K-1]$ in \textbf{[parallel]}}
    \For{$j$ in $[0\ldots n-1]$ in \textbf{[parallel]}}
        \State $i\gets 0$
        \For{$p$ in $[X,Y,Z]$ in \textbf{[parallel]}}
            \State $w\gets\text{pauli\_to\_weight}(p)$ 
            \For{$(g,\text{wire},\theta)$ in $U_{k,j}$}
                \State $w\gets\text{map}_{1q}(w,g(\theta))$
            \EndFor
            \State $(\text{LUT}_{1q})_{k,j,i}\gets w$
            \State $i\gets i+1$
        \EndFor
     \EndFor    
\EndFor
\State \Return $\text{LUT}_{1q}$
\end{algorithmic}
\end{algorithm}
\subsubsection{2-qubit gate}

For simplicity, we consider the CX gate, which acts on a control qubit (indexed by $c$) and a target qubit (indexed by $t$), modifying the $c^{\text{th}}$ and $t^{\text{th}}$ Pauli matrices in each Pauli string $P_{n,j}$. The CX gate transforms these Pauli matrices according to specific rules, such as: $XI\leftrightarrow XX, IY\leftrightarrow ZY, YI\leftrightarrow YX, IZ\leftrightarrow ZZ, XY\leftrightarrow YZ$ while leaving $II, IX, ZI$ and $ZX$ unchanged, or changing both Pauli and sign $(XZ\leftrightarrow -YY)$. For $m'$ CX gates, each Pauli string is updated according to these rules $m'$ times. These transformations are efficiently implemented using lookup table control ($\text{LUT}_c$) / target ($\text{LUT}_t)$ / sign ($\text{LUT}_{\text{sign}}$).

% , an example LUT can be shown as:

% \begin{align}
%    % & \text{LUT}_{c} = 
%    %  \begin{bmatrix}
%    %  0 & 0 & 3 & 3 \\
%    %  1 & 1 & 2 & 2 \\
%    %  2 & 2 & 1 & 1 \\
%    %  3 & 3 & 0 & 0
%    %  \end{bmatrix},
%    % \text{LUT}_{t} = 
%    % \begin{bmatrix}
%    %  0 & 1 & 2 & 3 \\
%    %  1 & 0 & 3 & 2 \\
%    %  1 & 0 & 3 & 2 \\
%    %  0 & 1 & 2 & 3
%    %  \end{bmatrix}, 
%     \text{LUT}_{\text{sign}} = 
%     \begin{bmatrix}
%     1 & 1 & 1 & 1 \\
%     1 & 1 & 1 & -1 \\
%     1 & 1 & -1 & 1 \\
%     1 & 1 & 1 & 1
%     \end{bmatrix}.
% \end{align}

The transformed Pauli matrices and sign for the control ($c^{\text{th}}$) and target ($t^{\text{th}}$) qubits of a CX gate are obtained from $\text{LUT}_{c/t/\text{sign}}$. Unlike single-qubit gates, which may produce weighted sums of Pauli strings, two-qubit gates like CX require the generator to be in a simple form before application. This operation is denoted as $\text{map}_{2q}:\mathbb{P}\times\{g_1,\ldots,g_m\}\rightarrow \mathbb{P}$, where $\mathbb{P}$ is a generator and $\{g_1,\ldots,g_m\}$ represents $m$ two-qubit gates, preserving the simple form of the output.

\section{Main algorithm}
\label{sec:algorithm}

\begin{algorithm}
\caption{Qimax v3. For v2, the indices tensor $\mathbb{I}$ is ignored.} 
\label{algo:Qimax}
\begin{algorithmic}
\Require instructions, $n$
\State $\{U_{k,j}\}, \{V_j\},K,K'\gets$ divide\_instruction(instructions)
\State order $\gets$ create\_chain($K,K'$); $\text{LUT}_{1q}\gets \text{create\_LUT}_{1q}(\{U_{j,i}\})$
\State $\Lambda^{(0)}, \mathbb{I}^{(0)}\gets [\text{create\_Z}_{j}(j,n)\;\textbf{for}\;j\;\textbf{in}\;[0,n]]; j\gets 0$
\For{order in orders}
    \State $k\gets j \div2$
    \If{order == 0} \Comment{$\{U_{k,j}\}$}
        \State $\Lambda,\mathbf{W}, \mathbb{I}\gets \text{sub}(\Lambda^{(t)},\mathbb{I}^{(t)},\text{LUT}_{1q}[k])$
        \State $\Lambda^{(t+1)}, \mathbb{I}^{(t+1)}\gets\text{flatten}(\Lambda,\mathbf{W}, \mathbb{I})$
    \Else \Comment{$\{V_{k}\}$}
        \For{$c,t$ in $V_k$} \Comment{Control and target wires}
            \State $\Lambda^{(t+1)}, \mathbb{I}^{(t+1)}\gets \text{map}_{2q}(\Lambda^{(k)}, \mathbb{I}^{(k)}, c, t)$
        \EndFor
    \EndIf     
    \State $j\gets j+1$
\EndFor
\State \Return $\{\Lambda^{(K+K')},\mathbb{I}^{(K+K')}\}$
\end{algorithmic}
\end{algorithm}

The initial generators $\{P_{n,j}^{(0)}\}=\{\mathbb{Z}_{n,j}\}$ are encoded in simple form $  \Lambda=\underbrace{[1,1,\ldots,1]}_{n},\quad\mathbb{I}=\underbrace{[[3], [3], \ldots, [3]]}_{n}$, as defined in~\ref{eq:generator}, where $\Lambda$ represents the coefficients (all 1) and $\mathbb{I}$ stores the Pauli $Z$ (index 3). In Qimax, the output of a single-qubit operator $U_k$ serves as the input to the next two-qubit operator $V_k$, and vice versa, until the final operator is applied. The function sub() (for single-qubit gate substitution) and $\text{map}_{2q}()$ (for two-qubit gate permutations) have low computational complexity due to their reliance on lookup tables. However, the primary computational bottleneck arises from the flatten() function, which converts generators from complex forms to simple forms.

\begin{align}\label{eq:flatten}
\text{flatten}:\hat{\mathbb{P}}_{n,j}^{(t)}\rightarrow\mathbb{P}_{n,j}^{(t)} = \sum_{k=0}^{n'-1}\left\{\left(\bigtimes_{j=0}^{n-1} \mathbf{w}^{(t)}_{j,k}    \right) \right\},
\end{align}

where $\bigtimes$ is Cartesian product notation. The function computes $n'$ combinations of weights from the tensor $\mathbf{w}^{(t)}$
, where $\mathbf{w}^{(t)}_{j,k}$ represents the weights for the $j^{\text{th}}$ qubit in the $k^{\text{th}}$ Pauli string. This involves iterating over all combinations of weights across qubits, with coefficients $\lambda_k$ derived from the product of weights. Since single-qubit gates preserve the identity Pauli $I$, the number of combinations is reduced by a factor of $(3/4)^{n_I}$ for a Pauli string $P_{n,k}$ containing $n_I$ Pauli $I$ operators, as $I$ contributes no additional terms. On GPU, this process is implemented as: (i) Each thread processes a subset of index combinations, (ii) coefficients are computed via parallel reduction (multiplication), and (iii) encoded indices are generated through base-4 accumulation.

The classical simulability of quantum circuits is commonly characterized by worst-case bounds. For state-vector simulation, the time and memory complexity scale as $T_{\mathrm{SV}}(n,m) = \mathcal{O}(m 2^n)$ and $M_{\mathrm{SV}}(n) = \mathcal{O}(2^n)$, respectively. For extended stabilizer simulation, the cost arises from stabilizer rank growth. 
Let $n_1(t)$ denote the number of Pauli strings after applying $t$ gates. In the worst case,  $n_1(t) \le 4^n$, the standard sequential cost can be expressed as:

\begin{align}
T_{\mathrm{seq}}(n,m) = O(m \bar{n}_1),
\end{align}

where $\bar{n}_1$ is the average stabilizer rank. In Qimax, grouping gates into $K$ single-qubit operators and $K'$ two-qubit operators yields

\begin{align}
T_{\mathrm{Qimax}}(n,m,P)
=
O\left(\frac{(K+K') \bar{n}_1}{P}\right)
+ O(F(n_1)),
\end{align}
where $F(n_1)$ denotes the flatten cost and $P$ is the number of GPU cores. Because typically $K+K' \ll m$, and $P \gg 1$ on modern GPUs $
T_{\mathrm{Qimax}} \ll T_{\mathrm{seq}}
\quad \text{whenever } 
P > \frac{m}{K+K'}.
$

This inequality formalizes the condition under which architectural reformulation shifts the practical simulability boundary. Importantly, Qimax does not alter the asymptotic worst-case upper bound $O(4^n)$ inherent to extended stabilizer formalism. 
However, by reducing synchronization depth and memory amplification, it expands the feasible simulation region $
\mathcal{R}_{\mathrm{feasible}}
=
\{ (n,m) \mid n_1(n,m) \le n_1^*(P, M_{\max}) \},
$ with $c$ is the memory cost per stabilizer generator per qubit:

\begin{align}
n_1^*(P, M_{\max})
=
\min\left(
\frac{P T_{\max}}{K+K'},
\frac{M_{\max}}{c n}
\right).
\end{align}

\section{Experiments}
\label{sec:experiments}

The three modes of Qimax are implemented in Python 3.10.11, using CuPy-cuda12x 13.4.0 \cite{cupy_learningsys2017} as the backend for GPU acceleration. Computationally intensive functions are executed in CUDA kernels with a block size of 256 and a grid size of $\lceil (\text{total elements} + 255) / 256 \rceil$, where total elements represent the number of tensor entries or tasks.

Qimax v1 implements the extended stabilizer formalism without encoding or GPU parallelization, serving as a CPU-based baseline with limited scalability. Qimax v2 uses a fixed-size weight tensor of dimensions $n\times n'\times n \times 4$. This structure simplifies memory allocation but consumes significant memory during the flatten() function due to the exponential number of Pauli string combinations. Qimax v3, however, employs a ragged weight tensor with a corresponding index tensor $\mathbf{i}$, reducing memory usage by ignoring zero-padding. This enables v3 to handle sparse representations efficiently, though it requires additional memory for storing indices.

\begin{figure}
    \centering
    \includegraphics[width=0.99\linewidth]{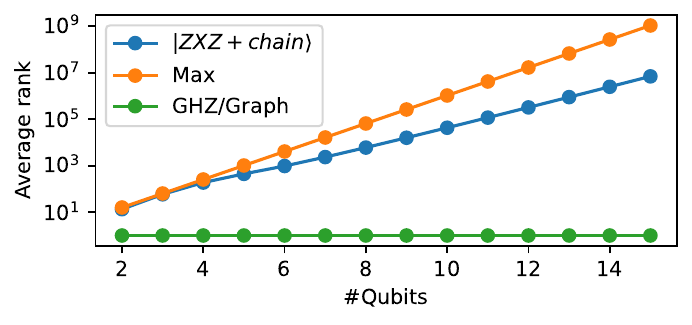}
    \vspace{-1cm}
    \caption{Average stabilizer rank (average rank) of the generators for different benchmark circuits. The stabilizer rank for the $|XYZ+\text{ chain}\rangle$ circuit grows exponentially with the number of qubits (approximately $\mathcal{O}(2^n)$), while the theoretical worst-case rank grows as $\mathcal{O}(4^n)$. In contrast, the GHZ and Graph circuits maintain a constant stabilizer rank of 1.}
    \label{fig:rank2}
\end{figure}

Comparable software includes PennyLane 0.41.0 \cite{bergholm2022pennylaneautomaticdifferentiationhybrid} and Qiskit 1.1.1 \cite{javadiabhari2024quantumcomputingqiskit}, both accelerated with NVIDIA’s cuQuantum SDK (lightning.gpu 0.41.0 and cusvaer\_simulator\_statevector 0.13.3, respectively). Experiments were conducted on an Intel i9-10940X CPU (3.30 GHz) and an NVIDIA RTX 4090 GPU. Execution time was measured from the initialization of stabilizer generators to the computation of final generators. Each experiment was run at least 10 times, and the average execution time was calculated. The benchmarked circuits include the GHZ circuit (multi-qubit entangled state), Graph circuit (graph-based quantum state), and $|XYZ+chain\rangle$, as shown in Figure~\ref{fig:operator} (b). The circuit structure consists of $L$ layers, with the XYZ component repeated $\#\text{Repeats}$ times within each layer. Experiments varied $\#\text{Repeats}$ from 100 to 100,000 to evaluate Qimax’s scalability. The \#Gates increases linearly with $\#\text{Repeats}$ and \#Qubits. The average stabilizer rank of generators, which impacts performance, is plotted in Figure~\ref{fig:rank2}; higher ranks correspond to lower performance in Qimax due to increased computational complexity.

\begin{figure*}[ht]
    \centering
    \includegraphics[width=0.99\linewidth]{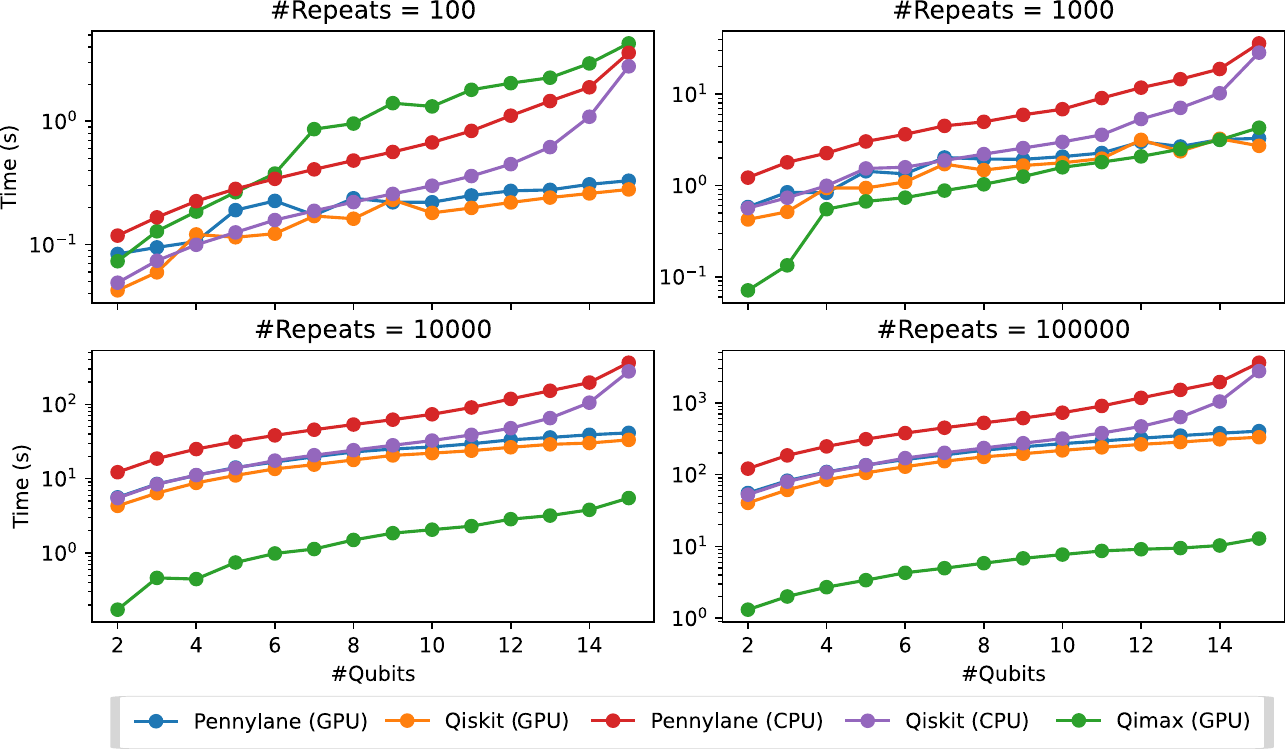}
    \caption{Execution time from different versions of PennyLane, Qiskit, and Qimax. The y-axis is plotted on logarithmic scale. The range of \#Qubits is 2 to 15.}
    \label{fig:times}
\end{figure*}

Qimax v1 shares properties with other stabilizer formalism packages, such as Stim \cite{stim}, enabling the simulation of many-qubit Clifford circuits in seconds, in this cases, GHZ and Graph circuits have the average stabilizer rank as 1. However, Qimax v1 is slower than Stim due to its support for non-Clifford gates, prioritizing functionality over performance. Qimax v3 demonstrates superior performance in experiments with large \#Repeats, as shown in Figure~\ref{fig:times}, outperforming GPU-accelerated Qiskit and PennyLane. At higher \#Repeats, Qimax v3’s performance advantage over these packages increases significantly. CPU-based versions of all packages exhibit the longest execution times, with Qiskit generally outperforming PennyLane. All packages exhibit exponential scaling with \#Qubits due to the state-vector size (for Qiskit and PennyLane) or the increase in gate count and average stabilizer rank (for Qimax).

% \begin{figure}
%     \centering
%     \includegraphics[width=0.70\linewidth]{images/ghz_graph.pdf}
%     \caption{The execution time for GHZ circuit and graph circuit on Qimax, these circuits are generated from MQT Bench \cite{quetschlich2023mqtbench}.}
%     \label{fig:ghz_circuit}
% \end{figure}

\section{Conclusion}
\label{sec:conclusion}

Qimax is the parallel version of stabilizer formalism using an encoding-decoding architecture. While operating on a multi-dimensional tensor speed-up by both software and hardware rather than the original stabilizers, Qimax can take advantage of the GPU. This approach offers a short execution time in case of a million gates, faster than GPU-based Qiskit and PennyLane, and even Qimax is implemented by Python. For high-stabilizer rank circuits, Qimax will be slower than the state-vector simulator due to the squared complexity ($\mathcal{O}(4^n)$ compared to $\mathcal{O}(2^n)$), same as other extended stabilizer formalisms. The next versions focus on optimizing multiple circuits' execution. However, Qimax's performance is sensitive to the stabilizer rank of the simulated circuits. For high-rank circuits, where $n'$ approaches the worst-case bound of $4^n$, Qimax's computational complexity scales as $\mathcal{O}(4^n)$, compared to $\mathcal{O}(2^n)$ for state-vector simulators. The performance advantage of Qimax is most pronounced in depth circuits and moderate stabilizer rank growth, where operator-level grouping and GPU tensor execution effectively amortize overhead. Therefore, Qimax is particularly well-suited for near-Clifford workloads with structured depth rather than arbitrary high-rank circuits. This limitation, common to other extended stabilizer formalism packages, arises from the exponential growth of Pauli string combinations during operations like the flatten() function. Despite this, Qimax v3 mitigates memory constraints through sparse tensor representations, enabling it to handle a broader range of circuits than its fixed-tensor counterpart (v2).

\section*{Acknowledgments}
This work was supported by JSPS KAKENHI Grant-in-Aid for Early-Career Scientists Grant Number JP26K23253, and partially supported by the VNUHCM - University of Information Technology’s Scientific Research Support Fund.
\bibliographystyle{IEEEtran}
\bibliography{references.bib}

\end{document}